\documentclass[superscriptaddress,showpacs,aps,twocolumn,amsmath,amssymb,prb]{revtex4}

\usepackage{graphicx} 
\usepackage{epstopdf} 
\usepackage{dcolumn}  
\usepackage{bm}       
\usepackage{txfonts}
\usepackage{textcomp}
\usepackage{verbatim} 
\usepackage{color}
\usepackage{array}
\definecolor{darkblue}{rgb}{0,0,0.6}

\newcommand{\papertitle}[2]{}

\newcommand{\bvec}[1]{{\text{\bf #1}}}

\DeclareMathOperator{\sign}{sign}

\newcommand{\bhline}[1]{\noalign{\hrule height #1}}

\newcommand{\red}{\color{red}}
\renewcommand{\red}{\relax}

\begin{document}

\title{Time-domain Brillouin scattering assisted by diffraction gratings}

\author{Osamu Matsuda}
\email{omatsuda@eng.hokudai.ac.jp}
\affiliation{Division of Applied Physics, Faculty of Engineering,
      Hokkaido University, 060-8628 Sapporo, Japan}

\author{Thomas Pezeril}
\affiliation{Institut Mol\'ecules et Mat\'eriaux du Mans, UMR CNRS 6283, Universit\'e du Maine, 72085 Le Mans, France}

\author{Ievgeniia Chaban}
\affiliation{Institut Mol\'ecules et Mat\'eriaux du Mans, UMR CNRS 6283, Universit\'e du Maine, 72085 Le Mans, France}
        
\author{Kentaro Fujita}
\affiliation{Division of Applied Physics, Faculty of Engineering,
      Hokkaido University, 060-8628 Sapporo, Japan}

\author{Vitalyi Gusev}
\email{Vitali.Goussev@univ-lemans.fr}
\affiliation{Laboratoire d'Acoustique de l'Universit\'{e} du Maine, UMR CNRS 6613, Universit\'{e} du Maine, Av. O. Messiaen, 72085 Le Mans, France}

\begin{abstract}
Absorption of ultrashort laser pulses in a metallic grating deposited
on a transparent sample launches coherent compression/dilatation
acoustic pulses in directions of different orders of acoustic
diffraction. Their propagation is detected by the delayed laser
pulses, which are also diffracted by the metallic grating, through the
measurement of the transient intensity change of the first order diffracted light. The
obtained data contain multiple frequency components which are
interpreted by considering all possible angles for the Brillouin
scattering of light achieved through the multiplexing of the
propagation directions of light and coherent sound by the metallic
grating. The emitted acoustic field can be equivalently presented as a
superposition of the plane inhomogeneous acoustic waves, which
constitute an acoustic diffraction grating for the probe light. Thus,
the obtained results can also be interpreted as a consequence of
probe light diffraction by both metallic and acoustic gratings. The
realized scheme of time-domain Brillouin scattering with metallic
grating operating in reflection mode provides access to acoustic
frequencies from the minimal to the maximal possible in a single
experimental configuration for the directions of probe light incidence
and scattered light detection. This is achieved by monitoring of the
backward and forward Brillouin scattering processes in
parallel. Applications include measurements of the acoustic
dispersion, simultaneous determination of sound velocity and optical
refractive index, and evaluation of the samples with a single
direction of possible optical access.

\medskip

\noindent
Key words: picosecond laser ultrasonics, time-resolved Brillouin scattering, inhomogeneous plane acoustic wave, acousto-optics
\end{abstract}

\pacs{43.35.Sx, 78.20.hc, 78.35.+c, 81.70.Cv, 43.38.Zp}

\maketitle
\section{Introduction} 

Picosecond acoustic interferometry (PAI) is a powerful
opto-acousto-optic technique for nondestructive and non-contact testing
of transparent materials at the
nanoscale.\cite{Thomsen:optcom1986,Grahn:ieeeq1989,Wright:optlet1991,Wright:jap1992,OHara:jap2001,Devos:apl2005,Mechri:apl2009,Steigerwald:apl2009,Yarotski:apl2012,Dehoux:softmat2012,IJChen:nlett2014,Nikitin:nrep2015,Dehoux:lsa2016}
First, using an
ultrashort pump laser pulse, a propagating picosecond coherent
acoustic pulse (CAP) is launched into the material. Second, partial
scattering of a continuously delayed in time ultrashort probe laser
pulse by the launched CAP is used to monitor the propagation of this
nanometers scale acoustic perturbation through the material. Weak
light pulses scattered by the CAP interfere at the photodetector with
the probe light pulses of significantly higher amplitude reflected
from various interfaces of the sample, such as the interfaces of the
tested material with air and with the optoacoustic transducer, for
example. The signal of transient optical reflectivity is proportional,
in leading order, to the product of the amplitudes of these two
scattered light fields. Thus, a heterodyning of a weak field against a
strong one is achieved. The detected signal of time-resolved optical
reflectivity in this so-called pump-probe scheme contains a sinusoidal
oscillating component whose physical origin is the Brillouin
scattering (BS) of the probe light by the CAP. The frequency of this
oscillation depends on the angle between the propagation directions of
the probe light and the coherent acoustic waves. It is precisely equal
to the shift in the frequency of the scattered light that would be
caused by the thermal phonons propagating in the same direction as the
CAP and could be resolved using optical spectrometers in classic
frequency-domain BS (FDBS) experiments.\cite{Fabelinskii:molecular,HayesLoudon:scattering,Dil:rpp1982,Polian:jrs2003}  That is why the PAI is
also often called time-domain Brillouin scattering (TDBS).

An important limitation of TDBS in comparison with FDBS is that a
significantly narrower part of the acoustic spectra is accessible by TDBS.
In FDBS, by varying the angle between the directions
in which probe light is incident and in which scattered light is
detected, the direction of the thermal phonons for testing can be
selected.\cite{Thomsen:optcom1986,Grahn:ieeeq1989,Wright:jap1992}
Because thermal phonons are available in all directions, this
approach is very flexible providing the opportunity to significantly
vary the angle between the directions of the probe light and the
phonon wave vectors. Thermal phonons of highest frequency are
accessible in the so-called back-scattering
configuration,\cite{Wright:jap1992} when the probe light is scattered
by the counter-propagating or co-propagating phonons (annihilation or
creation of phonons, respectively). Thermal phonons of lowest
frequency are detectable in the forward scattering
configuration/geometry when the probe light and the phonons are
propagating along nearly orthogonal directions, as in the so-called
platelet configuration/geometry.\cite{Polian:jrs2003} In TDBS the
situation is very different. In common TDBS experiments the lateral
dimensions of the optoacoustic generators are controlled by the
size of the laser pump focus and typically significantly exceed the
spatial lengths of the CAP emitted by them in the materials. Thus, the
diffraction length of the emitted CAP in typical experiments
significantly exceeds its attenuation length, while the direction of
the CAP is quasi-perpendicular to the sample
surface illuminated by the pump laser pulse, and fixed. Then the only
possibility to vary the frequency of the tested phonon is to change
the direction of the probe light propagation relative to the fixed
direction of CAP propagation. However, most of the TDBS applications
are for the diagnostics of either thin coatings/multilayers deposited
on the bulk samples or thin plates/membranes with an optical access to
the launched CAP only through the surfaces normal to CAP propagation
direction. The maximum angle of the probe light transmitted through
the air/material surface relative to the direction of CAP propagation
is theoretically $\theta_i^\text{max}=\arcsin(1/n)$, where $n$ is the
refractive index of the material at the probe wavelength $\lambda$, when in air the light is skimming
along the surface. Thus for $n>\sqrt{2}$, the angle is smaller than 45$^\circ$. In
the resulting backscattering-type configuration the frequency of the
phonon tested by the TDBS is
\cite{Lin:jap1993,Cote:rsi2005,Lomonosov:acsnano2012}
\begin{equation}
  f_B=2vn\sqrt{1-\sin^2(\theta_i)}\bigg/\lambda=2vn\sqrt{1-\sin^2(\theta)/n^2}\bigg/\lambda,
  \label{eq:fbpp}
\end{equation}
where $v$ is the longitudinal sound velocity of the material and
$\theta$ is the angle of incidence in air. Thus for $n>\sqrt{2}$, even
the lowest frequency theoretically accessible by TDBS,
$f_B^\text{min}$, is close to the maximum one,
$f_B^\text{min}>f_B^\text{max}/\sqrt{2}\simeq 0.71f_B^\text{max}$. For
more convenient but still large angle of incidence, $\theta=60^\circ$,
the estimate is $f_B^\text{min}>\sqrt{2/3}f_B^\text{max}\simeq
0.82f_B^\text{max}$.  Thus, the tunability of the Brillouin frequency
expressed in Eq.~\eqref{eq:fbpp} is limited by refraction at the
air/sample interface.  This indicates the limitations in application of
the TDBS for revealing and identifying the frequency dispersion of the
material properties such as, for example, sound velocity and
attenuation. Another limitation of the typical TDBS scheme is that, in
accordance with Eq.~\eqref{eq:fbpp}, when fixing the external angle
of probe incidence and measuring the Brillouin frequency (BF), $f_B$,
we get information on the combination of two material parameters ($v$
and $n$). Then to determine them independently the TDBS measurements
should be conducted at least at two different angles
$\theta$,\cite{Cote:rsi2005,Tomoda:apl2007,Lomonosov:acsnano2012} while in
FDBS there exists an experimental forward scattering configuration,
which is called platelet configuration,\cite{Polian:jrs2003}
providing the opportunity to measure sound velocity by a single
measurement without determining the optical refractive index.

To get access in the TDBS experiments to larger angles of BS,
including those of forward-type scattering, and, as a consequence, to
broaden the spectrum of the detectable/accessible phonons and to
measure the sound velocity within a single mutual orientations of the
optical excitation and detection, we propose to use gratings
consisting of periodically arranged pump light absorbing parallel
bars. The bars could be metallic, for example, as in our experiments
(Fig.~\ref{fig:setup}). When the grating with a period $p$ is applied
for the generation of the CAPs instead of a metallic thin film,
typically used for this purpose, the coherent acoustic waves will be
emitted not only normally to the plane of the gratings but also in all
directions corresponding to the possible diffraction of the acoustic
wave by this grating, thus, multiplexing the propagation directions of
CAPs in the sample. Additionally the grating can diffract both
transmitted and reflected probe light, thus, multiplexing the
propagation directions of the probe laser pulses inside the sample. Both
these factors should potentially lead to an increase of the maximum
angles between the propagation directions of the coherent sound
and of the probe light, and give access for TDBS to forward-type photon
scattering processes. Moreover, application of the grating should
provide an opportunity to monitor simultaneously the same acoustic
mode, for example longitudinal, at different frequencies.

\section{Earlier experiments with metallic grating diffracting probe
  light in transmission mode}

After realizing these theoretical predictions experimentally in the
schema presented in Fig.~\ref{fig:setup}, we have found that a part
of them could have been confirmed just through the dedicated analysis
of the experimental results published much earlier in
Ref. \onlinecite{Lin:jap1993}. In this publication the pump-probe
optical schema was for the first time successfully applied to reveal
the elastic motions of the metallic grating deposited on the samples
surface (gold rods on fused silica, the same combination of materials
as in our experiments). The principle difference in comparison with
our optical experiments is in conducting pumping and probing of the
samples from the grating side of the sample.  {\red The experimental setup
was similar to Fig.~\ref{fig:setup} but the sample was placed upside down:
both pump and probe light comes from the front side (with grating) of
the sample.}  The title of this
publication, ``Study of vibrational modes of gold nanostructures by
picosecond ultrasonics'', and its abstract, both emphasize the
successful identification of the low-lying frequencies in the
transient reflectivity spectrum with the normal modes of the nanorods
coupled to the substrate. Because of this fact and the years passed we
forgot that in Ref.~\onlinecite{Lin:jap1993} three additional
high-lying frequencies (modes I--III in Fig.~10 of
Ref.~\onlinecite{Lin:jap1993}) had been detected and reproduced by
the solutions of the theoretically formulated problem, although the
origin of not all of them had been understood even qualitatively. The
origin of mode I was identified with the backscattering-type process
described in Eq.~\eqref{eq:fbpp} without any influence of the
grating on it. The origin of the mode II was related with the
influence of the grating on the probe light field only, i.e., without
accounting for the difference in the directivity patterns of the CAPs
emitted by the metallic grating and a metallic thin film. Finally the
origin of mode III was not understood. Based on our proposal that
grating directs both the probe light and the generated acoustic waves
in different orders of the diffraction, the interpretation of the
physical origin of the modes II and III is straightforward. If in the
coordinate system presented in Fig.~\ref{fig:setup} the wave vector
of the probe photon incident from air on the sample is given by
$\bvec{k}_i=(k_x,0,k_z)=(k_x,0,\sqrt{k^2-k_x^2})$, where
$k=|\bvec{k}|$, then in transmission from air into the sample the
probe field is diffracted by the grating in multiple directions
defined by the wave vectors
$\bvec{k}_i=(k_x+m_iq,0,\sqrt{k_i^2-(k_x+m_iq)^2})$. Here $q=2\pi/p$
is the grating wave number, $k_i=nk$ is the wave number of the probe
photon in the sample, while $m_i=0,\pm1,\pm2,\dots$ indicates the
order of the diffraction peak. The probe light field backscattered by
the phonons should have the propagation directions described by
$\bvec{k}_s=(k_x+m_sq,0,-\sqrt{k_s^2-(k_x+m_sq)^2})$,
$m_s=0,\pm1,\pm2,\dots$. Only the light propagating along these
directions, when transmitted from the sample into the air, could be
diffracted by the grating in the detection direction, given in
Ref.~\onlinecite{Lin:jap1993} by
$\bvec{k}=(k_x,0,-\sqrt{k^2-k_x^2})$. Note that the indexes $i$ and
$s$ are introduced for the photon incident on the phonon and scattered
by the phonon, respectively. These photons are propagative in the
limited number of the diffraction orders defined by
$k_{i,s}>|k_x+m_{i,s}q|$ and evanescent in the rest. The wave vector
of the acoustic phonon participating in the BS is given by the law of
the momentum conservation,\cite{Fabelinskii:molecular,HayesLoudon:scattering,Dil:rpp1982,Polian:jrs2003}
$\bvec{k}_B=\pm(\bvec{k}_s-\bvec{k}_i)$ where the plus sign corresponds
to absorption of the acoustic phonon (anti-Stokes) and the minus sign
to its emission (Stokes). Then modulus of the wave vector of the
coherent acoustic phonon which has participated in the BS is
\begin{equation}
 \begin{split}
   k_B&=\frac{2\pi f_B}{v}\\
   &=\Biggl\{[(m_s-m_i)q]^2\\
   &\qquad+\left[d_s\sqrt{k_i^2-(k_x+m_sq)^2}-d_i\sqrt{k_i^2-(k_x+m_iq)^2}\right]^2\Biggr\}^{1/2},
 \end{split}
 \label{eq:fbdf}
\end{equation}
Here the difference in the frequencies of the incident and scattered
photons is neglected, $k_s=k_i$, as usually, while the parameters
$d_{s,i}=\pm1$ are introduced by us to account for the variety of
possible directions of incident and scattered light, in the general
case. In the experiments in Ref.~\onlinecite{Lin:jap1993},
$d_i=-d_s=1$ corresponds to monitoring of the backscattered light
only, because the forward scattered light does not return to the
detection region. The proposed Eq.~\eqref{eq:fbdf} provides the
frequency of the mode I as in Eq.~\eqref{eq:fbpp} when
$m_s=m_i=0$. The proposed Eq.~\ref{eq:fbdf} reproduces Eq. (14) from
Ref.~\onlinecite{Lin:jap1993} when $m_s=m_i=m$ and, thus, reveals the
physical sense of the parameter $m$ in
Ref.~\onlinecite{Lin:jap1993}. The derived condition also confirms the
suggestion in Ref.~\onlinecite{Lin:jap1993} that in the mode II the
light is scattered by the plane acoustic front propagating in the
direction normal to the surface. In fact, the projection of the phonon
wave vector along the surface (along the $x$ axis) is proportional to
the difference between the projections of the photons wave vectors,
$(m_s-m_i)q$, and, thus, for the revealed $m_s=m_i=m$ it is equal to
zero. The phonon scattering light in mode II propagates normally to
the surface.  Moreover the experimentally observed mode II corresponds
to $|m|=1, \sign(m)=-\sign(k_x)$. So the origin of the mode II is the
scattering by the CAP of the light directed by the metallic grating in
such diffraction order whose direction is the closest to the CAP
propagation direction. Finally, Eq.~\eqref{eq:fbdf} attributes the
origin of the mode III to the following two degenerate in frequency
processes. In the first one the metallic grating directs the incident
light like in mode II, i.e. in the diffraction order closest to the
surface normal, but then the coherent acoustic wave scatters light in
such direction towards the front surface of the sample, from which it
can be detected without additional diffraction ($|m_i|=1,
\sign(m_i)=-\sign(k_x), m_s=0$). In the second one the acoustic wave
backscatters non-diffracted (zeroth order) probe light in the
diffraction order of light closest to the surface normal, from which
it is returned to the detection direction by the optical grating in
transmission from the sample into the air ($m_i=0, |m_s|=1,
\sign(m_s)=-\sign(k_x)$). In both these processes the acoustic waves
are not just reflecting the incident light like the mirrors but are
modifying the direction of the scattered light relative to one
predicted by the Snell's law. One can say that the acoustics waves are
diffracting the incident light and are functioning as diffracting
gratings with the wave number $q$. It is natural in the following to
use for the acoustic field, generated in the sample by pump laser
pulses incident on the metallic grating, the term acoustic grating
because it is periodic along the $x$ axis. Note, that this terminology
was used already for example for the description of the acoustic waves
generated by laser induced gratings, i.e., by the intensity
interference patterns that can be created by two light beams incident
on the sample surface at angles $\theta$ and
$-\theta$.\cite{Gusev:apl2009,Gusev:jap2010,Kouyate:jap2011,Kouyate:josab2016}
It was demonstrated that the acoustic field emitted by the laser
gratings can be decomposed into the so-called plane inhomogeneous
acoustic
modes,\cite{Gusev:apl2009,Gusev:jap2010,Kouyate:jap2011,Kouyate:josab2016}
i.e., the acoustic waves with plane phase fronts parallel to the
laser-intensity grating but modulated in amplitude with the pattern of
the laser intensity grating. Thus these modes can be naturally called
acoustic gratings. In the experiments in
Ref.~\onlinecite{Lin:jap1993} and in our experiments presented
below in Section III (Fig.~\ref{fig:setup}) acoustic gratings are
generated by the metallic gratings. Their ability not just to
reflect/transmit but to diffract the incident light is due to their
amplitude periodic modulation. Using the suggested terminology all the
high-lying frequency modes detected in
Ref.~\onlinecite{Lin:jap1993} can be explained by processes
involving only two diffractions of probe light by the gratings, i.e.,
either two diffractions by metallic grating (mode II) or one diffraction by the
metallic grating plus one diffraction by the acoustic
grating (mode III). Although in these experiments only the backscattering-type
processes were observed and only those involving first diffraction
orders, they are in favor of our proposal formulated in the
Introduction that in the experiments with gratings multiple
frequencies corresponding to BS processes with different angles
between sound and light propagation directions can be detected
simultaneously.

\section{Experiments with metallic grating diffracting probe light in reflection mode}

\begin{figure}
\centering
\includegraphics[width=0.8\linewidth]{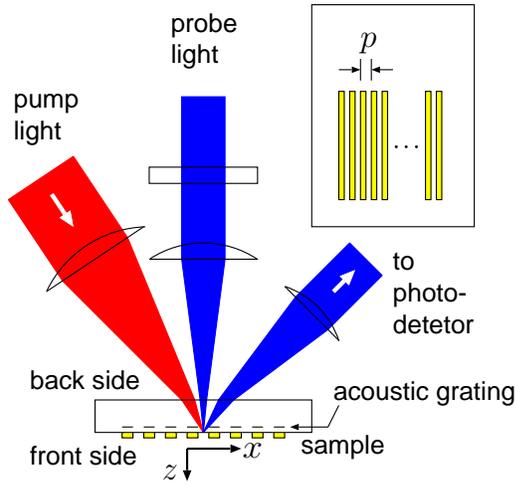}
\caption{The optical setup for the measurement. Both pump and probe
  beams are incident from the backside (without grating) of the
  sample. Signal is detected on the first order diffraction emitted
  from the backside of the sample. Insert: Schematic diagram of the
  periodic gold stripes formed on a fused silica substrate. The
  nominal period $p$ is 587.0 nm or 479.2 nm, corresponding to first
  order optical diffraction of the probe light in air at around 45 and
  60 degrees, respectively.  {\red The grating spacing is significantly
  exagerated: in reality several gold strips are accommodated within the laser
  spot so that the light is well diffracted.}}
\label{fig:setup}
\end{figure}

In our TDBS experiments conducted from the opposite side of the sample
(Fig.~\ref{fig:setup}) than in Ref.~\onlinecite{Lin:jap1993}, we have
additionally detected the BFs corresponding to forward scattering
processes, to the processes involving three diffractions of probe
light (two by metallic grating and one by acoustic grating) and also
to the processes involving light from the higher diffraction orders
than the first order. The gold gratings are made by the electron beam
lithography and lift off technique on a fused silica substrate of
thickness 1 mm. Insert in Fig.~\ref{fig:setup} shows a schematic
structure of the sample. Two samples with different nominal grating
periods, $p=587.0$ nm and $479.2$ nm, are prepared. We refer the
former sample as “45deg”, whereas the latter as “60deg”, because
they are prepared to have the first order diffraction peaks near these
directions for the reflected probe light incident normally at the
grating. The designed width of each rod is half the period, but the
actual width is somewhat larger. The thickness of the gold rods is
about 50 nm. A 2 nm thick Cr layer is formed between the SiO$_2$
substrate and the Au film to improve the adhesion.

A standard laser picosecond setup is used. A mode-locked Ti:sapphire
laser with a regenerative amplifier is used as the light source. The
pulse width is $\sim$100 fs and the repetition frequency is 260
kHz. The fundamental light pulses with the central wavelength 800 nm
are focused to the grating structure from the back side (the side
without grating) of the sample (Fig.~\ref{fig:setup}). The pulse
energy is 80 nJ/pulse and the diameter of the focused region is 100
$\mu$m, covering nearly completely the rectangular grating area, which
lateral dimensions are similar (100 $\mu$m by 100 $\mu$m). The
absorption of pump laser pulses in the metallic grating generates
acoustics waves propagating in the sample along different orders of
acoustic diffraction, as it is described in the
Introduction. Alternatively the launched acoustic field could be
viewed as inhomogeneous plane waves or acoustic grating propagating
normally to the sample
surface.\cite{Gusev:apl2009,Gusev:jap2010,Kouyate:jap2011,Kouyate:josab2016}
The acoustic grating, propagating normal to the metallic grating is
the result of the interference of the acoustic waves propagating along
positive and negative orders of the acoustic diffraction.

\begin{figure}
{\flushleft (a)\par}
  
\includegraphics[width=\linewidth]{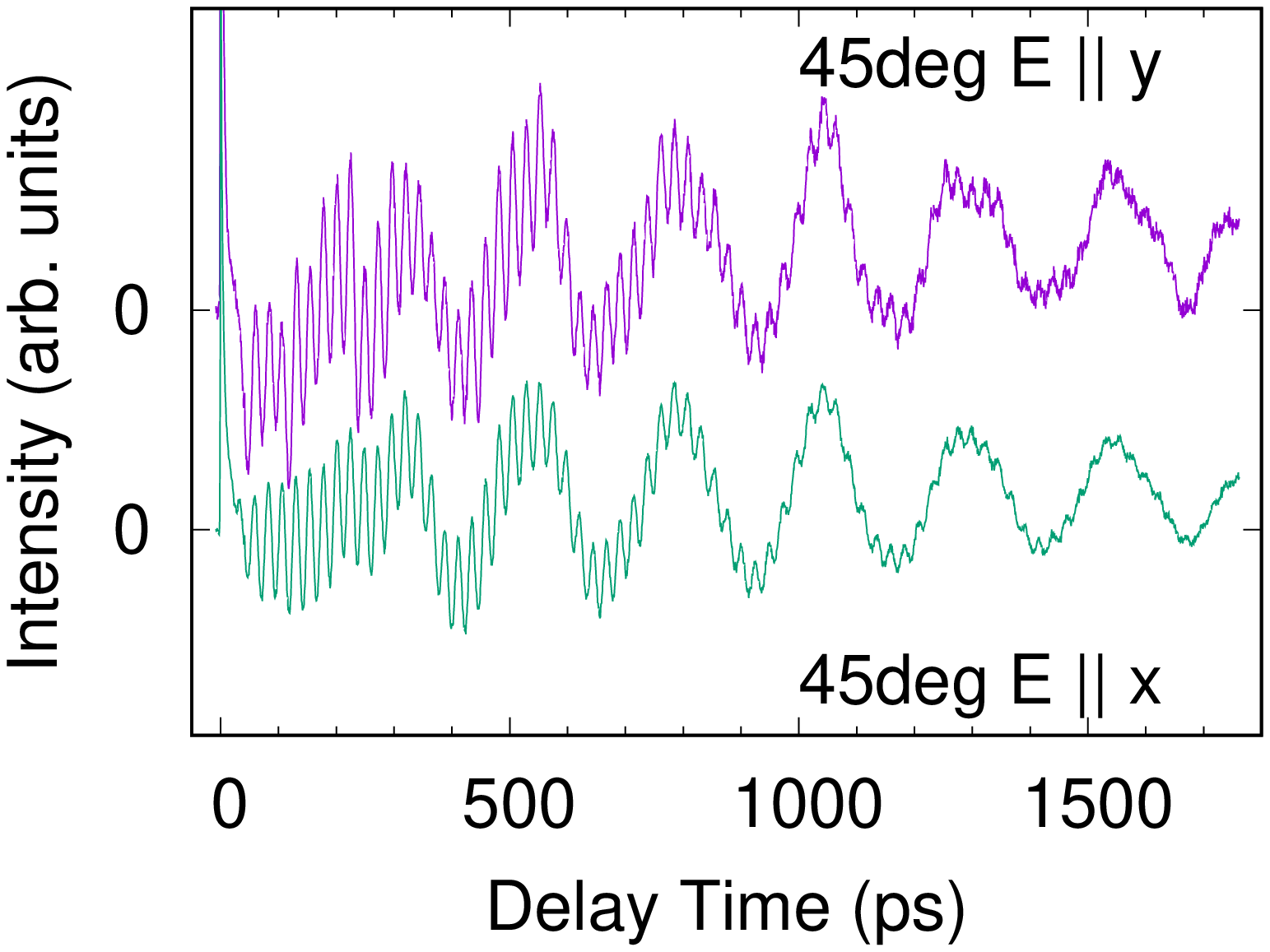}

{\flushleft (b)\par}

\includegraphics[width=\linewidth]{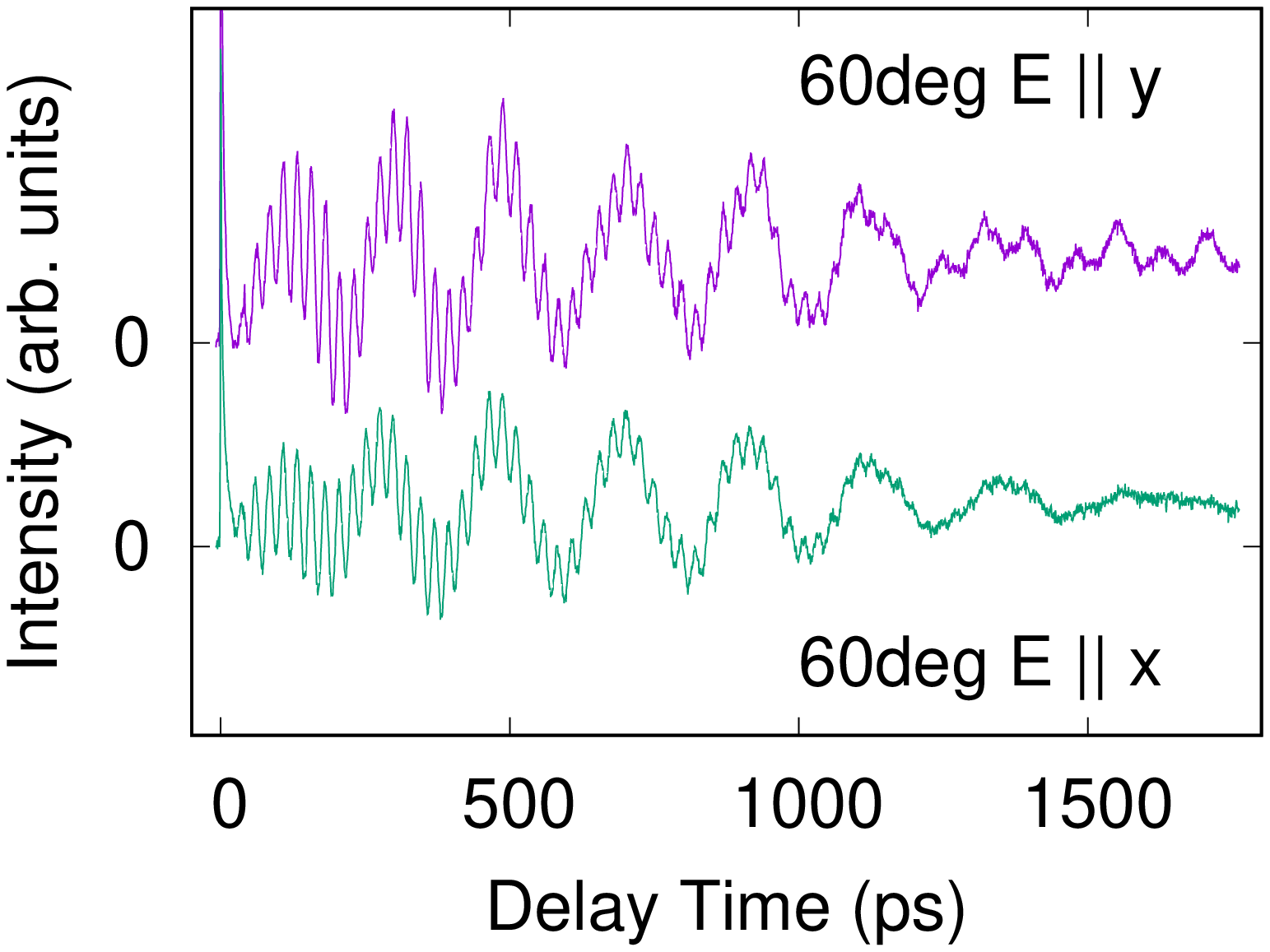}

\caption{
The transient variation of the intensity of the first order diffracted
light as a function of the delay time between the pump and probe light
pulse arrival to the sample. (a) for 45deg sample, and (b) for 60 deg
sample. The polarization of the incident probe light is also noted in
the figure.
}
\label{fig:rawdata}
\end{figure}

The frequency doubled light pulses with the central wavelength 400 nm
are focused to the grating structure from the back side with the
normal incidence. The pulse energy is 4 nJ/pulse and the diameter of
the focused region is 10 $\mu$m. The probe light scattered by the
complete structure in the first order of diffraction is fed in the
photo detector to reveal the modulation of the light intensity caused
by the acoustic waves. (Fig. \ref{fig:setup}). The polarization of the
incident probe light is chosen as in parallel with the grating period
($x$ axis, $\bvec{E}\parallel\bvec{x}$) or in parallel with the gold
bars ($y$ axis, $\bvec{E}\parallel\bvec{y}$). Figure \ref{fig:rawdata}
shows the raw data of the transient variation of the intensity of the
diffracted light for 45deg and 60deg samples with two probe
polarizations $\bvec{E}\parallel\bvec{x}$ and
$\bvec{E}\parallel\bvec{y}$. The delay time is the time between the
pump and probe light pulse arrival to the sample.


\begin{figure}
{\flushleft (a)\par}
  
\includegraphics[width=\linewidth]{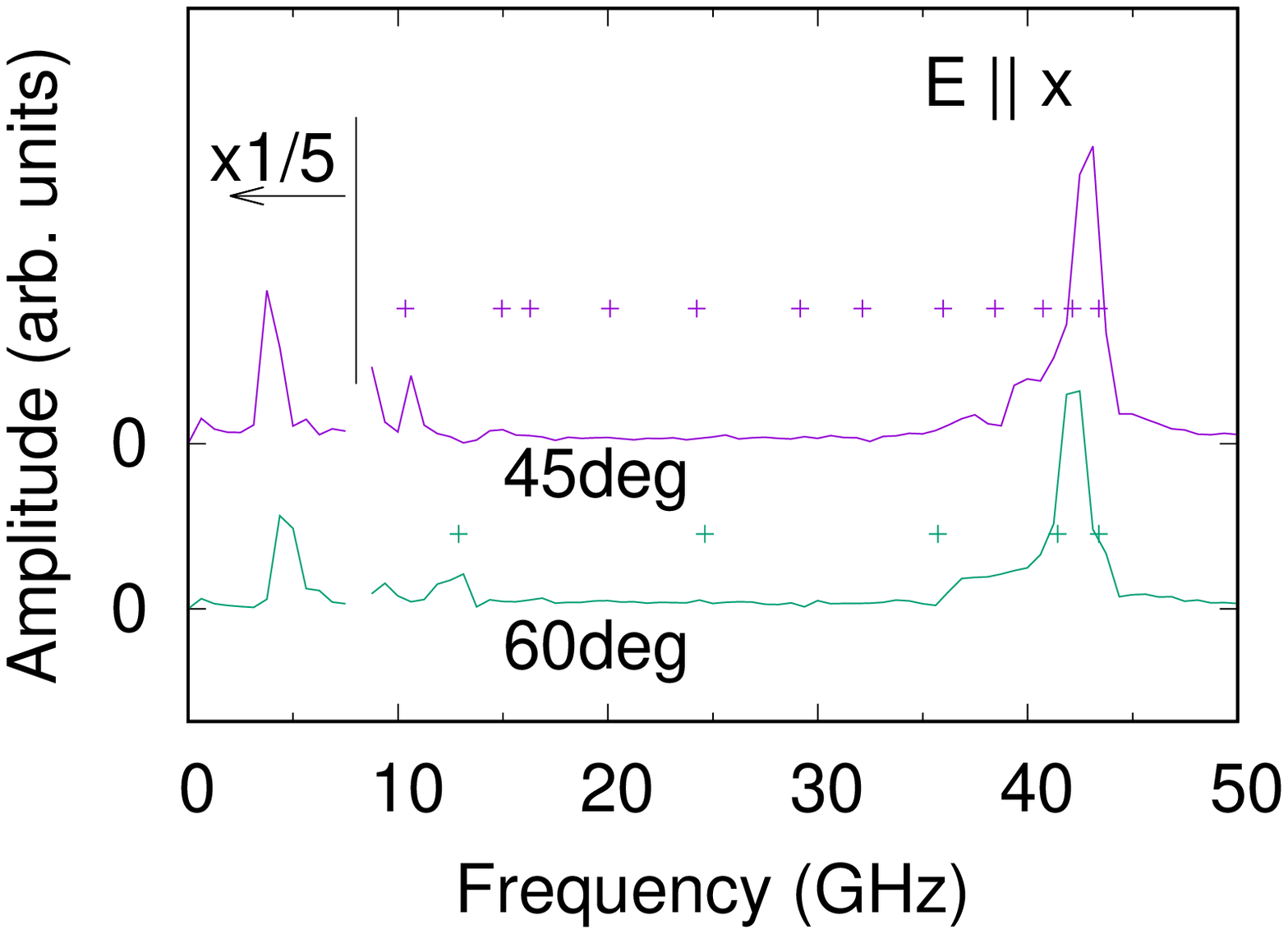}

{\flushleft (b)\par}

\includegraphics[width=\linewidth]{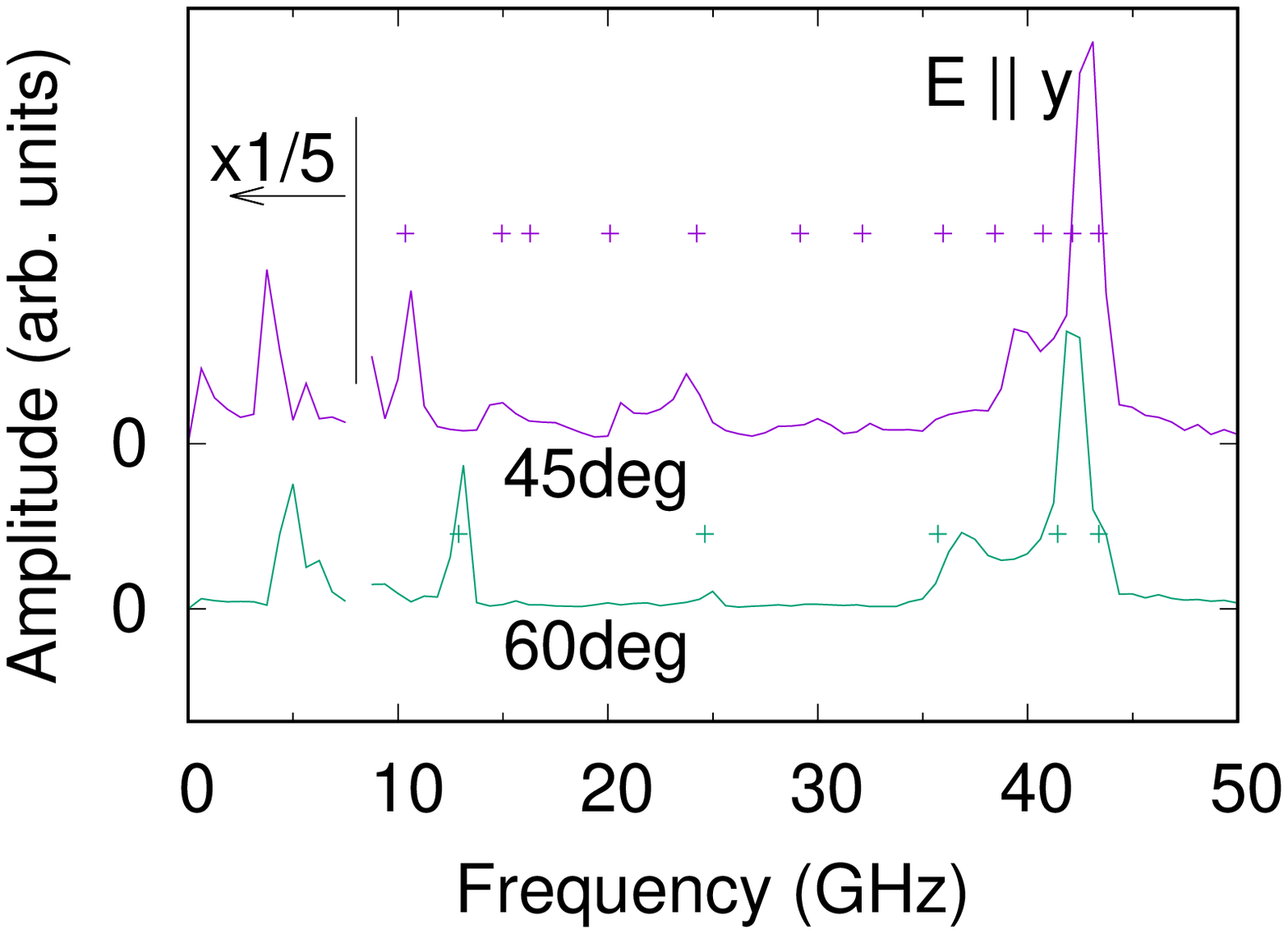}

\caption{The norm of Fourier amplitude of the data shown in
  Fig. 2. (a) for $\bvec{E}\parallel\bvec{x}$, and (b) for
  $\bvec{E}\parallel\bvec{y}$. The $+$ symbols denote the
  theoretically calculated frequencies of possible Brillouin
  scattering configurations. For convenience of presentation the
  amplitudes in the low-lying part of the spectra ($<8$ GHz) are 5
  times diminished.  }
\label{fig:ftdata}
\end{figure}

Each transient curve consists of two contributions: a slow oscillating
components with a period around 200 ps, and a fast oscillating
components with a period around 20 ps and resembles the curves
presented in Fig.~2 of Ref.~\onlinecite{Lin:jap1993}. This indicates a
possible splitting of the photo-induced motion of the sample into
low-lying frequency modes and high-lying frequency modes as in
Ref.~\onlinecite{Lin:jap1993}. To get further understanding, the
obtained temporal signal is Fourier transformed with respect to the
delay time. Figure \ref{fig:ftdata} shows the norm of Fourier
amplitude as a function of frequency.

In the low-lying part of the spectrum below 8 GHz, where the
amplitudes in Fig.~\ref{fig:ftdata} are intentionally attenuated, we
detect, similar to Ref.~\onlinecite{Lin:jap1993}, up to three
different modes. Following Ref.~\onlinecite{Lin:jap1993} we attribute
them to the oscillations of the gold rods on the sample
surface. Although these oscillations are beyond of our interest here,
it was straightforward to associate the strongest of the low-lying
modes (at 3.8 GHz and 5.0 GHz in the 45deg and 60deg samples,
respectively) with the L (longitudinal) mode defined in
Ref.~\onlinecite{Lin:jap1993}. This was achieved using the results of
numerical calculations of the resonance frequency of the rods with
different widths and thicknesses presented in Fig. 8 of
Ref.~\onlinecite{Lin:jap1993} and accounting for the fact that
frequency of the strongest mode in our experiments scales
approximately inverse proportional to its width.


High-lying parts of the observed spectra in our experiments are always
containing larger number of frequency peaks in comparison with
Ref.~\onlinecite{Lin:jap1993}. For example in 45deg sample in case of
$\bvec{E}\parallel\bvec{y}$ probe polarization the number of the
frequencies that can be identified is twelve (see
Fig.~\ref{fig:ftdata} (b) and Table \ref{tbl:freq}), i.e., four times
larger than that in Ref.~\onlinecite{Lin:jap1993}. All the
experimentally observed modes in the high-lying part of the spectrum
can be attributed to particular BS processes theoretically
contributing to Eq.~\eqref{eq:fbdf}, when accounting for the
differences between our experimental configuration and samples and
those in Ref.~\onlinecite{Lin:jap1993}. The first and the most
important advantage of the optical probing the sample from the back
side (Fig.~\ref{fig:setup}) consists in the additional opportunity to
monitor forward-type BS processes by the TDBS. This opportunity is
related to the fact that in our experiments the metallic grating acts
on the probe light in the reflection mode while in the optical schema
of Ref.~\onlinecite{Lin:jap1993}, i.e., {\red when the sample in
Fig.~\ref{fig:setup} is placed upside down and probed from the front side,} it acts
on the probe light in the transmission mode only. This can be
qualitatively understood considering four types of the probe light
scattering sequences which are contributing to the reflection of light
by the sample in the direction of the detector. Accounting for the
fact that the scattering of light by the acoustic grating is much
weaker than by the metallic one, only the sequences with a single
scattering by the acoustic grating should be considered. The first and
the simplest sequence is the diffraction of the probe light normally
incident on the acoustic grating in reflection/backscattering mode
towards the detector. This process corresponds to $k_i=0$,
$d_s=-d_i=-1$, $m_i=0$, $m_s=1$ in Eq.~\eqref{eq:fbdf}. We remind here
that the parameters $m_{i,s}$ numerate the diffraction orders, while
$d_{i,s}$ fix the direction of the probe light propagation along the z
axis. The detection direction is $d_s=-1$, $m_s=1$ in our
experiments. The second sequence including backscattering of light by
the acoustic grating includes: a) the transmission of the probe light
through the acoustic grating without diffraction, b) reflection or
diffraction of light by metallic grating in the different orders of
the diffraction, c) reflection/backscattering of light by the acoustic
grating in the different orders of the diffraction (the process with
$k_i=0$, $d_s=-d_i=1$ and arbitrary $m_{i,s}$ in Eq.~\eqref{eq:fbdf}),
d) reflection or diffraction of the light by the metallic grating in
the direction of its detection. Two other sequences include forward
scattering of light by the acoustic grating
($\sign(d_s)=\sign(d_i)$). The third sequence consists in: a) forward
scattering of the probe light normally incident on the acoustic
grating in the different orders of the diffraction (the process with
$k_i=0$, $d_s=d_i=1$, $m_i=0$ and arbitrary $m_s$ in
Eq.~\eqref{eq:fbdf}) and b) reflection or diffraction of the light by
the metallic grating in the direction of its detection. The fourth
sequence includes: a) the transmission of the probe light through the
acoustic grating without scattering, b) reflection or diffraction of
light by metallic grating in the different orders of the diffraction
and c) forward scattering/transmission of the light by the acoustic
grating in the direction of the first order of the diffraction (the
process with $k_i=0$, $d_s=d_i=-1$, arbitrary $m_i$ and $m_s=1$ in
Eq.~\eqref{eq:fbdf}). It is worth noting here that in our 1 mm thick
samples in order to be detected the light scattered from region of the
acousto-optic interaction towards the back surface should propagate in
the direction of the detection. In fact, because of the large
thickness of the sample in comparison with the lateral dimension of
the metallic and acoustic gratings, the light scattered in the other
diffraction orders after the reflection from the back surface does not
incident on the metallic and acoustic gratings and, thus, it cannot be
later redirected by them to the detector.

\begin{figure}
\centering
\includegraphics[width=\linewidth]{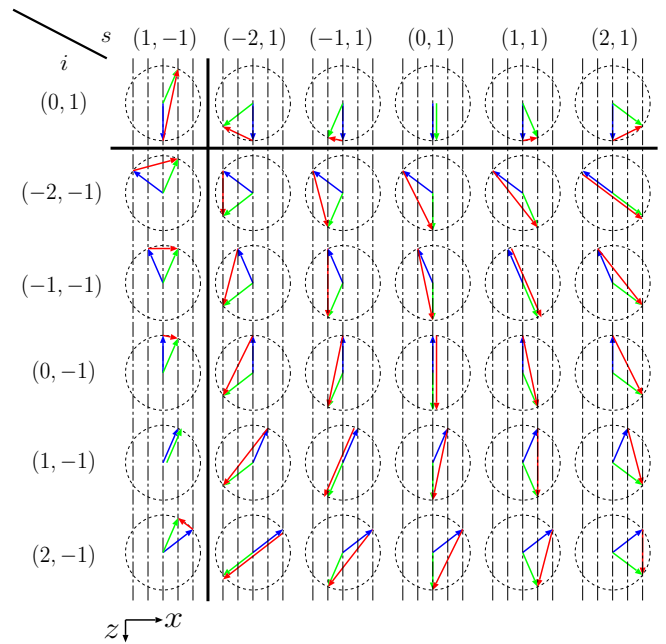}
\caption{The possible Brillouin scattering configurations for the
  normal incident probe light with the wave vector $(0, 0, k_i)$ and
  the detection at the probe light along the first diffraction order
  with the wave vector $(q, 0, -((k_i)^2- q^2)^{1/2})$, both in the
  substrate. The wave vectors of photons before the scattering,
  $\bvec{k}_i$, are shown in blue arrows. The wave vectors of the
  photons after the scattering, $\bvec{k}_s$, are shown in green
  arrows. The wave vectors of the phonons, $\bvec{k}_B$, are shown in
  red arrows. The $x$ and $z$ axes are shown at the bottom. In the
  left column the state of the photon incident on the acoustic grating
  is identified by the diffraction order $m_i$ and the direction of
  its propagation along the $z$ axis, $d_i=\pm1$, in the form
  $(m_i,d_i)$. In the upper row the state of the photon scattered by
  the acoustic grating is identified by the diffraction order $m_s$
  and the direction of its propagation along the $z$ axis, $d_s=\pm1$,
  in the form $(m_s,d_s)$. The distance between the dash-dot vertical
  lines is equal to the wave number of the gratings, $q$. The black
  lines separate the forward scattering processes (in the lower left
  and the upper right corners) from the backward scattering
  processes. The vertical and horizontal continuous lines indicate the
  distinction of forward and backward scattering processes: The
  upper-left and lower-right processes involve the backward
  scattering, whereas the upper-right and lower-left processes involve
  the forward scattering.}
\label{fig:scattering}
\end{figure}

We have found that for the identification of all BFs detected in all
four conducted experiments it is sufficient to account only for the
propagating incident and scattered light fields in the processes
defined by Eq. (2), although only the light directed to the detector
should be obligatory propagating (because of the large distance
between the region of its diffractions/scatterings and the back
surface of the sample). Although in the second of the above described
sequences both incident and scattered light fields could be
potentially evanescent, while in the third and the forth sequences the
evanescent could be the scattered and the incident light,
respectively, it was sufficient for us to account (in addition to the
zeroth order of the diffraction) only for the first and the second
orders of the diffracted light in the 45deg sample and for the first
order of the diffracted light in the 60deg sample. With the refractive
index of the fused silica in our sample $n=1.47$ the light in all
other diffraction orders is evanescent
($k_{i,s}<|k_x+m_{i,s}q|$). Thus it was possible to explain all the
observed BFs only by the BS processes with the momentum conservation
diagrams presented in Fig.~\ref{fig:scattering}. In this Figure and
later on the state of the incident photon (left column) is identified
by $(m_i,d_i)$, while the state of the scattered photon (upper row) by
$(m_s,d_s)$. Blue, green and red arrows are presenting the wave
vectors of the incident photon, scattered photon and the phonon,
respectively.

The wave vector diagrams for the Brillouin scattering processes,
corresponding to the four probe light transmission/reflection
sequences described above, are presented in the upper left, lower
right, upper-right and lower-left parts of Fig.~\ref{fig:scattering},
respectively. They are separated by continuous black lines. Equation
\eqref{eq:fbdf} (for $k_x=0$) provided opportunity to calculate the
frequencies of all scattering processes depictured in
Fig.~\ref{fig:scattering}. The calculated BFs are presented in Table
\ref{tbl:freq} together with the experimentally registered. For
completeness we listed in the left column in Table \ref{tbl:freq} all
the cases of degeneracy, i.e., when different scattering
configurations in space actually correspond to the same angle between
the propagation directions of the incident photon and the phonon. The
notation ``NA'' in Table \ref{tbl:freq} marks the processes involving
the photons from the second diffraction order in the 60deg sample,
which are evanescent. It was not required to account for these
processes to explain the totality of our experimental observations.

\begin{table}[tb]
  \caption{Calculated (calc.) and experimental (exp.) frequencies of
    phonons involved in each Brillouin scattering configuration
    (config.). The value in the parentheses indicates the
    corresponding peak is substantially broadened. The dash symbol
    indicates that the corresponding peak is not observed (exp.). The
    lines with empty values indicate that the mentioned configuration
    is degenerated to the one with values below. NA means the
    corresponding scattering mode involves evanescent photons. Thick
    horizontal line in the middle separates the upper part, where
    there the data can be explained by the scattering processes
    involving probe light of the zeroth and the first orders of
    diffraction, from the lower part, where all the processes
    involve additionally the photons from the second diffraction
    order. Frequency unit: GHz.}
  \begin{tabular}{ccccccc}
    \hline
    \hline
    config. & 45deg & & & 60deg &  \\
            & calc. & exp. & exp. & calc. & exp. & exp. \\
            & & $(\bvec{E}\parallel\bvec{y})$ & $(\bvec{E}\parallel\bvec{x})$ & & $(\bvec{E}\parallel\bvec{y})$ & $(\bvec{E}\parallel\bvec{x})$ \\
    \hline
    $(0,-1)\rightarrow(1,-1)$ \\
    $(0,1)\rightarrow(\pm1,1)$ & 10.35 & 10.6 & 10.6 & 12.90 & 13.1 & 12.9 \\
    \hline
    $(-1,-1)\rightarrow(1,-1)$ & 20.10 & 20.6 & --    & 24.62 & 25.0 & -- \\
    \hline
    $(\pm1,-1)\rightarrow(\pm1,1)$ & 38.44 & 39.4 & 39.8 & 35.72 & 36.9 & 36.8 \\
    \hline
    $(0,1)\rightarrow(1,-1)$ \\
    $(0,-1)\rightarrow(\pm1,1)$ \\
    $(\pm1,-1)\rightarrow(0,1)$ & 42.13 & 42.5 & 43.0 & 41.42 & 41.9 & 41.9 \\
    \hline
    $(\pm2,-1)\rightarrow(\mp2,1)$ \\
    $(\pm1,-1)\rightarrow(\mp1,1)$ \\
    $(0,-1)\rightarrow(0,1)$ & 43.38 & 43.3 & 43.1 & 43.38 & 43.4 & 43.4 \\
    \bhline{1.8pt}
    $(2,-1)\rightarrow(1,-1)$ & 14.95 & 15.0 & -- & NA & NA & NA \\
    \hline
    $(0,1)\rightarrow(\pm2,1)$ & 24.24 & 23.7 & -- & NA & NA & NA \\
    \hline
    $(\pm2,-1)\rightarrow(\pm1,1)$ \\
    $(\pm1,-1)\rightarrow(\pm2,1)$ & 29.16 & 30.0 & -- & NA & NA & NA \\
    \hline
    $(-2,-1)\rightarrow(1,-1)$ & 32.12 & 32.5 & -- & NA & NA & NA \\
    \hline
    $(0,-1)\rightarrow(\pm2,1)$ \\
    $(\pm2,-1)\rightarrow(0,1)$ & 35.98 & (36.5 ) & (37.4) & NA & NA & NA \\
    \hline
    $(\pm2,-1)\rightarrow(\pm2,1)$ & 16.30 & -- & -- & NA & NA & NA \\
    \hline
    $(\pm1,-1)\rightarrow(\mp2,1)$ \\
    $(\pm2,-1)\rightarrow(\mp1,1)$ & 40.72 & -- & -- & NA & NA & NA \\
    \hline
  \end{tabular}
  \label{tbl:freq}
\end{table}

The theoretical estimates for Table \ref{tbl:freq} were done with the
following parameters of fused silica $n=1.47$ and $v=5968$
m/s.\cite{Handbook:CRC2003} The calculated values are also marked with
$+$ symbol in Fig.~\ref{fig:ftdata}. The correspondence between the
theoretically predicted and the measured Brillouin frequencies is
remarkable.

\section{Discussion}

We have already noted above that the larger number of Brillouin
frequencies accessible in our experiments in comparison with those in
Ref.~\onlinecite{Lin:jap1993} is due to the ability to monitor forward
scattering processes in the configuration presented in Fig. 1. Another
reason is in the larger number of the diffraction orders with
propagating probe light in our 45deg sample. We have checked that in
Ref.~\onlinecite{Lin:jap1993} in the samples with 400 nm and 600 nm
periodicity of the metallic grating only the photons in the zeroth and
the first diffraction orders could be propagative. In our 45deg sample
the photons are propagative additionally in the second diffraction
order importantly increasing the number of the efficient BS
configurations as it could be appreciated from the lower part of the
Table \ref{tbl:freq}, where the theoretical interpretation of the
detected BFs necessities the participation in the BS the photons from
the second diffraction order. The fact, the photons in the second
diffraction order are propagating in 45deg sample and are evanescent
in the 60deg samples also explains our experimental observations that
larger number of the BFs were detected in the first of this samples
(see Fig.~\ref{fig:ftdata} and Table \ref{tbl:freq}).

It was expected that the propagation of the probe light in the higher
diffraction orders would require the participation in the BS processes
of the phonons launched in the higher orders of diffraction, i.e.,
higher orders of light diffraction by the acoustic grating, in
comparison with the earlier experiments. This is confirmed by our
experimental results presented in Table \ref{tbl:freq} and explained
by the diagrams in Fig.~\ref{fig:scattering}. Only in one of the
processes, $(-1,-1)\rightarrow(1,-1)$, involving the first diffraction
orders of light (dominating in the upper part of the Table
\ref{tbl:freq} above the thick horizontal line in its center) the
participation of the phonon of the second diffraction order, i.e.,
with the projection of the phonon wave vector on the x axis equal to
$|m_s-m_i|q=2q$, is required, for explaining the particular
experimentally detected frequency. Note, that two other processes in
the upper part of the Table \ref{tbl:freq},
$(\pm1,-1)\rightarrow(\mp1,1)$ and $(\pm2,-1)\rightarrow(\mp2,1)$, due
to phonons emitted by the optoacoustic generator in the second and the
fourth diffraction orders, respectively, are possible. However, these
processes are degenerate in BF with the process
$(0,-1)\rightarrow(0,1)$, taking place without diffraction but just
through the reflection of light by the plane compression/dilatation
acoustic wave. The latter process takes place without diffracted
acoustic waves and is dominating over two other processes contributing
to the same BF. Consequently, accounting for the processes
$(\pm1,-1)\rightarrow(\mp1,1)$ and $(\pm2,-1)\rightarrow(\mp2,1)$ is
not necessary for the explanation of our experimental
observations. They are presented in Table \ref{tbl:freq} only for the
sake of completeness. Thus, the experimental BFs from the upper part
of Table \ref{tbl:freq} can be explained by the processes involving
photons of the zeroth and the first diffraction orders only. At the
same time the participation of the phonons from the second and even
from the third diffraction orders (see for example
$(-2,-1)\rightarrow(1,-1)$) is largely required in the processes
with the photons from the second diffraction order in the lower part
of Table \ref{tbl:freq}.

To explain all the available experimental data, we do need to account
for the phonon which is neither propagating nor decaying along the $z$
axis. The experimental frequencies of 20.6 GHz and 25.0 GHz, observed
only in the experiment with $\bvec{E}\parallel\bvec{y}$ polarization
of probe light in 45deg and 60 deg samples, respectively, can be
currently associated only with the scattering of light by longitudinal
phonons skimming along the sample surface. The corresponding
scattering process is $(-1,-1)\rightarrow(1,-1)$.

The observation of the BFs only when using probe light polarized along
the metallic rods, i.e., $\bvec{E}\parallel\bvec{y}$, is rather common
to our experiments (see Fig.~\ref{fig:ftdata} and Table
\ref{tbl:freq}). We attribute these observations to higher
reflection/diffraction of $\bvec{E}\parallel\bvec{y}$ polarized light
in comparison with the light polarized along the direction of grating
periodicity, i.e., $\bvec{E}\parallel\bvec{x}$, both by metallic and
acoustic gratings. On the one hand, this hypothesis is consistent with
the known applications of metal gratings as birefringent light
polarizers.\cite{LZhou:optlet2005,LZhang:apb2006} On the other hand,
the scattering of $\bvec{E}\parallel\bvec{x}$ polarized light may have
lower efficiency than that of $\bvec{E}\parallel\bvec{y}$ when the
propagation direction of the scattered light is nearly perpendicular
to the direction of the polarization which is induced by the incident
light and the acoustic waves.\cite{Kouyate:josab2016,BornWolf:1999}

Our experimental results confirm that the application of the
diffraction grating in the TDBS experiments provides opportunity to
overcome some limitations of the TDBS technique discussed in the
Introduction. First the proposed experimental scheme provides
opportunity to detect simultaneously multiple Brillouin frequencies
from the highest possible BF in the backscattering configuration,
$f_B^\text{max}=43.38$ GHz, to the lowest possible BF in the forward
scattering configuration, $f_B^\text{min}=10.6$ GHz (see the first and
the last line in the upper part of Table \ref{tbl:freq}). This
significantly broadens the frequency band of the TDBS from
$f_B^\text{min}\approx 0.7f_B^\text{max}$ (see Introduction) to
$f_B^\text{min}\approx 0.25f_B^\text{max}$. The highest BF detected in
our experiments does not depend on the grating, while the lowest
detected frequency is controlled by the grating period as it can be
appreciated from the diagrams in Fig.~\ref{fig:scattering} of the
several degenerate processes providing access to this frequency. It
could be additionally diminished by the increasing the period $p$ of
the grating. The comparison of the lowest BFs detected in our two
samples (first line in Table \ref{tbl:freq}) supports this
expectation. Thus the frequency range accessed by TDBS could be
additionally broadened by the dedicated preparation of the grating.

Our experiments demonstrate that the TDBS measurement in a single
experimental configuration with diffraction grating, i.e., without any
modification of the directions for optical pumping and probing of the
sample, is sufficient to extract refractive index, n, and sound
velocity, $v$, of the material. For this purpose any two of the
experimentally detected frequencies, which in Eq.~\eqref{eq:fbdf}
depend both on $n$ and $v$, could be used. It is also possible to
determine the refractive index and sound speed by fitting
simultaneously larger number the measured frequencies in order to
increase statistically the reliability in their determination, if
required.\cite{Lomonosov:acsnano2012} Moreover the detection of
the BF corresponding to the forward scattering process of light by the
skimming longitudinal wave, $(-1,-1)\rightarrow(1,-1)$, provides
opportunity to determine sound velocity without knowledge of the
refractive index. In this process $k_B=2\pi f_B/v=2q$ (see
Fig.~\ref{fig:scattering}) and the determination of the sound velocity
requires just the knowledge of the grating period, $v=pf_B/2$. If our
experiments were accomplished differently, with probe light incident
on the sample at angle $\theta$ to the sample normal and the scattered
light detected in the direction of mirror-type reflection, i.e., as in
Ref.~\onlinecite{Lin:jap1993} but from the backside of the sample, the
processes of forward scattering of probe light by the skimming
longitudinal wave would be still accessible by the TDBS. For example,
if the angle $\theta$ is chosen such that $k_x=k\sin\theta=-mq/2$,
then the acoustic grating composed of acoustic waves skimming along
the surface could scatter the probe light incident on it from the back
surface of the sample forward in the ``symmetrically'' propagating
light with $k_x=mq/2$ by transmitting it in the $m$-th diffraction
order of the grating. Later in the reflection from the metallic
grating this probe light could be directed to the detector. Thus the
momentum conservation law for scattering of the probe light by the
phonons skimming along the surface reads $k_B=|m|q=2|k_x|$. The first
of these equalities can be used to evaluate the sound velocity from
the measured BF without knowledge of the optical refractive
index. However, there is still a drastic difference in this approach
to evaluate sound velocity by TDBS with how it could be accomplished
in FDBS. In FDBS for any angle $2\theta$ between the direction of probe
incidence and probe detection the thermal phonon with the wave vector
necessary for the BS exists. For example, in the platelet geometry of
the FDBS, which has the same momentum conservation diagram as the
geometry providing access to skimming phonons by the TDBS and also
provides opportunity to measure sound velocity without knowledge of
the optical refraction angle, the thermal phonons with the required
$\bvec{k}_B$ exist for any $2\theta$. In the TDBS the coherent
skimming phonons can be generated only with wave vectors, which are
multiples of the metallic grating wave number, $q$. Thus, to access
these phonons the angle of incident should be chosen to satisfy the
equality $\sin\theta=mq/(2k)$. In both of our samples such
measurements could be potentially accomplished only at two different
angles of light incidence.

Finally, it is worth mentioning that in our experiments the number of
the detected forward scattering processes is smaller than the backward
scattering ones (see Fig.~\ref{fig:scattering}). This is related to
the fact that our scheme allows all possible forward scattering
processes to be accessed in the second sequence of the probe light
scattering from those described earlier. In this sequence in the
acousto-optic interaction both the light incident on the acoustic
grating and the scattered probe light can be of the arbitrary
diffractions orders, while in the third and the fourth scattering
sequences for the detection of the forward scattering processes only
one of these light fields could be of an arbitrary diffraction
order. This asymmetry in the TDBS scheme could be potentially
corrected in the thinner experimental samples with two diffraction
gratings deposited on the opposite sides when multiple reflections of
the probe light between the surfaces of the sample are becoming
important while the second grating (deposited on the back surface of
the sample in Fig.~\ref{fig:setup}) could, in particular, transmit in
the detection direction in air the probe light incident on it from the
sample side along an arbitrary order of diffraction. With the
diffraction gratings on both sides of the sample all BS processes
possible in the sample, both for forward and backward scattering,
could potentially involve incident and scattered light in arbitrary
diffraction orders.

\section{Conclusions}

We have performed picosecond ultrasonic interferometry (time-domain
Brillouin scattering (TDBS)) measurements in transparent samples
with metallic gratings. The pump light pulse absorbed in the metallic
grating structure generates acoustic gratings (inhomogeneous plane
compression/dilatation acoustic waves) in the substrate. The
propagation of the acoustic waves is monitored by delayed probe
light pulses. By detecting the modulation of the probe light intensity
in the first order diffracted beam, we observed in the time domain
Brillouin oscillations with rich frequency spectra. The obtained
results are explained by a theoretical model which takes into account
all possible configurations of probe light scattering/diffraction by
these acoustic gratings, including those where the light itself is
diffracted by the metallic grating either before its scattering by the
phonons or after this scattering, or both. The agreement between the
experimental positions of the Brillouin frequencies and the calculated
ones is excellent.

The theory revealed two reasons for the increased number of the BS
processes that could be monitored in the scheme of TDBS proposed by us
in comparison with the earlier reported experiments with metal
gratings.\cite{Lin:jap1993} Our scheme provided for the first time
access by TDBS to the forward scattering processes of light by the
coherent sound, while one of our gratings provided propagating probe
light in higher diffractions orders than in
Ref.~\onlinecite{Lin:jap1993}. Access to forward scattering processes
importantly broadens the range of frequencies accessible by the
TDBS. This fact in combination with the opportunities to monitor
multiple different BS processes/frequencies simultaneously would be
advantageous in studying dispersion of sound velocity and sound
attenuation in the materials. It is worth noting here that the
applications of usual TDBS schemes for studying acoustic wave
attenuation are documented for a variety of the
media.\cite{Lin:jap1991,Chen:pmb1994,Emery:apl2006,Devos:prb2008,Decremps:prl2008,Pezeril:prl2009,Ayrinhac:prb2011,Maznev:optlet2011,Klieber:jcp2013,Maehara:jjap2014,Danworaphong:apl2015}
The opportunity to monitor in a single measurement the acoustic
phonons propagating in different directions could be attractive for
revealing the elastic/inelastic anisotropy of the materials, including
one that could be caused by non-isotropic loading or by the residual
stress. For the studies of the anisotropy it is also extremely
advantageous that gratings, as demonstrated by our experiments, can
simultaneously launch phonons, detectable by TDBS, in the complete
diapason of the angles, i.e., from 0 degrees to 90 degrees relative to
their surface. Advantageous in certain practical applications with
limited optical access to the
samples\cite{Nikitin:nrep2015,Decremps:prl2008,Kuriakose:usonics2016}
would be provided by the opportunity to monitor by the TDBS with
grating multiple BS processes/frequencies even with incident and the
detected/scattered light propagating collinearly (for example in the
case of the probe light incident normally on the grating and the
detection of either reflected or transmitted light in the directions
also normal to the grating surfaces).

Finally some of the functionalities of the proposed scheme could be
achieved by replacing the metal gratings by laser-induced gratings
that could be generated by the interference pattern of two light beams
propagating at an angle. Such gratings can launch acoustic gratings in
the sample\cite{Gusev:apl2009,Gusev:jap2010,Kouyate:jap2011}
and also to diffract probe light,\cite{Kouyate:josab2016}
although much less efficiently than the metallic ones. The advantage
of laser gratings is in the perspective of the non-contact and
non-invasive diagnostics of the samples by TDBS. The drawback is in
more technically elaborated optical scheme.

\bigskip

\begin{acknowledgments}
The reported research was conducted in the frame of the project
PLUSDIL supported by ANR, under contract ANR-12-BS09-0031.  O.M. is
partially supported by the Acoustic HUB of R\'{e}gion des Pays de
La Loire in France, by a Grant-in-Aid for Scientific Research from
Japan Society for the Promotion of Science, and by a research grant
from the Murata Science Foundation.  We are grateful to
Prof. Humphrey J. Maris for the comments.  We would like to thank OPEN
FACILITY (Hokkaido University Sousei Hall) for the sample fabrication.
\end{acknowledgments}

\bibliography{matsuda}



\end {document}